\documentclass[11pt,a4paper,twoside]{article}
\usepackage[utf8]{inputenc} 
\usepackage[T2A]{fontenc} 
\usepackage[english]{babel} 
\usepackage{times} 
\usepackage{geometry} 
\usepackage{multicol} 
\usepackage{authblk} 
\usepackage{graphicx} 
\usepackage{amsmath} 
\usepackage{float} 
\usepackage{amssymb} 
\usepackage{fancyhdr} 

\date{} 
\title{STUDY OF SPATIAL INHOMOGENEITIES OF COSMIC RAYS IN A SYNTHETIC TURBULENT MAGNETIC FIELD}

\author[1]{P.~K. Batrakov}
\author[2]{V.~O. Yurovsky}
\author[3]{I.~A. Kudryashov}

\affil[1]{P.N. Lebedev Physical Institute of the Russian Academy of Sciences (LPI). \texttt{pavel.batrakov100@mail.ru}}
\affil[2]{Moscow State University named after M.V. Lomonosov, Faculty of Physics. \texttt{yrovskyvladimir@gmail.com}}
\affil[3]{Moscow State University named after M.V. Lomonosov, Skobeltsyn Institute of Nuclear Physics (SINP MSU). \texttt{ilya.kudryashov.85@gmail.com}}

\begin{document}
\maketitle
\pagestyle{fancy}
\fancyhf{} 
\fancyfoot[R]{\thepage} 
\renewcommand{\headrulewidth}{0pt} 
\renewcommand{\footrulewidth}{0pt} 

\fancypagestyle{firstpage}{
    \fancyhf{}
    \fancyfoot[R]{\thepage} 
    \renewcommand{\headrulewidth}{0pt}
    \renewcommand{\footrulewidth}{0pt}
}

\begin{center}
    \textbf{Abstract}
\end{center}

The paper presents a theoretical model describing the full power spectra of synchrotron radiation generated by relativistic electrons in a turbulent magnetic field. Using the theoretical model, numerical calculations of the complete power spectra of synchrotron radiation were performed for a turbulent field generated by the harmonic method. Additionally, a model sky map was constructed, demonstrating the structure of the spatial inhomogeneity of the synchrotron radiation power distribution as seen by an observer.

\vspace{1em} 
    \textbf{Keywords:} cosmic rays, synchrotron radiation, turbulent magnetic field.

\vspace{1em} 

\begin{multicols}{2} 
\section{INTRODUCTION}

Galactic cosmic rays (GCRs) are a flux of high-energy charged particles with energies ranging from approximately 10$^6$ eV to 10$^{18}$ eV. A key feature of GCRs is the non-thermal distribution of their energies. GCRs gain energy through acceleration in specific astrophysical processes of plasma and electromagnetic nature [1]. The study of the dynamics of galactic cosmic rays is an indirect method of investigating the galaxy's magnetic fields, alongside well-known methods such as polarization plane rotation and the Zeeman effect.

The main goal of this work is to analyze the behavior of relativistic particles within the structure of a simulated turbulent magnetic field. The outcome of the study includes the construction of synchrotron radiation spectra of the simulated particles and the generation of a model map showing the distribution of the total radiation power in the inhomogeneous model magnetic field.

\section{The Nature of Galactic Cosmic Rays}
More than 90 $\%$ of cosmic ray particles are hydrogen atoms, about 7 $\%$ are helium, and less than 1 $\%$ is contributed by heavier elements, while the contribution of the electron-positron component is $10^{-3}$$\%$. It is important to note that a complete and comprehensive theory of cosmic ray origin is currently lacking. A finished theory of cosmic ray origin should explain the main characteristics of cosmic rays: the power-law form of the energy spectrum, the magnitude of energy density, the chemical composition of primary and secondary cosmic rays, including data on the fluxes of antiprotons, electrons, positrons, gamma quanta, the practical constancy over time of cosmic ray intensity, and their very weak anisotropy.

According to current understanding, the electron-positron component in cosmic rays is generated in the processes of pulsar activity and supernova explosions (primary cosmic rays). The secondary electron-positron component arises from the interactions of the proton component of cosmic rays with galactic gas, resulting in the reaction that produces electrons and positrons as one of the decay products in the chain  p + X $\longrightarrow$ $\pi$$^{\pm}$$ \longrightarrow $$\mu$$^{\pm}$ $ \longrightarrow$ e$^{\pm}$. Despite the low intensity of electron and positron transport in cosmic rays, studying this component is very important for constructing self-consistent models of cosmic ray acceleration and transport [2].

\section{Synchrotron spectrum in the GCR}
Relativistic electrons in a magnetic field propagate along a helical trajectory along the field lines. The cyclotron frequency $\omega_{b}$ of the particles is expressed as the ratio of the magnetic field strength \(\mathbf{B}\) to the relativistic gamma factor \(\gamma = \frac{E}{m_{0}c^{2}}\), while the radius of the orbit (gyroradius) \(\mathbf{R}\) is calculated taking into account the velocity of high-energy charged particles \(\mathbf{v}_{e}\), their mass \(m_{e}\), and the speed of light \(c\):

\begin{equation}
\label{one} 
\omega_{b} = \frac{e\textbf B}{m_{e}c\gamma}.
\end{equation}

\begin{equation}
\label{one} 
\textbf R= \frac{c\textbf v_{e}}{e\textbf B}\gamma.
\end{equation}

In the process of such transport, high-energy electrons experience energy losses due to magnetobremsstrahlung (synchrotron) radiation. The spectrum of synchrotron radiation spans from radio emission to soft X-ray emission. Synchrotron radiation is always directed along the trajectory of high-energy particles at an opening angle \(\theta = \arcsin\frac{1}{\gamma}\).

The spectral distribution of the total power of synchrotron radiation \(\mathbf{P(\omega)}\) is expressed using the MacDonald function \(K_{\frac{5}{3}}(z)\) (the modified Bessel function of the second kind), where the critical frequency \(\omega_{c}\) depends on the pitch angle \(\alpha\) [3,4]:

\begin{equation}
\label{one} 
\omega_{c}=\frac{2}{3}\omega_{b}\gamma^{3}\sin(\alpha)=\frac{3eB}{2m_{e}c}\gamma^{2}\sin(\alpha)
\end{equation}

\begin{equation}
\label{one} 
P(\omega)=\frac{\sqrt{3}}{2\pi}\frac{e^{3}B\sin(\alpha)}{mc^{2}}F(\frac{\omega}{\omega_{c}}) ,
\end{equation}

where

\begin{equation}
\label{one} 
F(\frac{\omega}{\omega_{c}})=\frac{\omega}{\omega_{c}}\int_{\frac{\omega}{\omega_{c}}}^{\infty}K_{\frac{5}{3}}(z)dz
\end{equation}

The spectral distribution of the total power (in all directions) of radiation from a charged particle moving in a magnetic field is presented in (Fig. 1), where the maximum occurs at \(\frac{\omega}{\omega_{c}} = 0.29\).

\begin{figure*}
\centering{\includegraphics[scale=0.8]{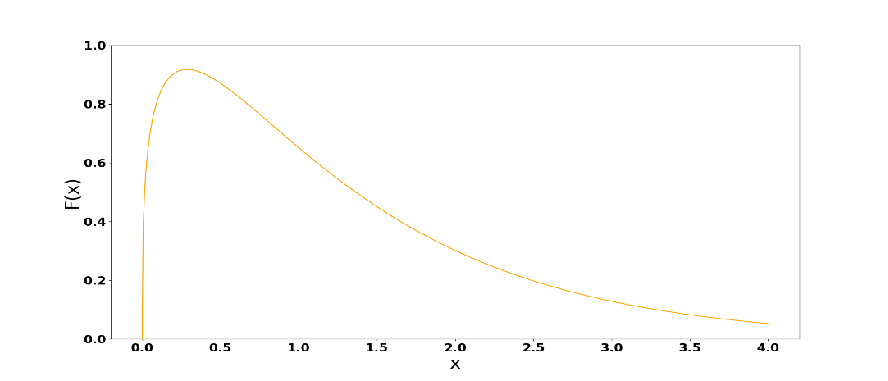}}
\caption{The values of the function \(F(x)\), obtained through numerical calculation, where \(x\) is the ratio \(\frac{\omega}{\omega_{c}}\), and \(F(x)\) is described by Equation 5.}
 \label{G-photo}
\end{figure*}

\section{Numerical Experiment}
The main objective of the study is to create a model that predicts the spatial spectrum of inhomogeneities in the synchrotron radiation of electrons in the GC. The first part of the work involved developing a magnetic field model that simulates the actual relationship between the turbulent and regular components of the magnetic field in the galaxy. The implementation of the magnetic field was carried out using the harmonic method. This method allows for the generation of the magnetic field \(\mathbf{B}\) at arbitrary points in space with position vectors \(\mathbf{r}\) through the superposition of \(N\) plane waves with random amplitudes \(A_{n}\), phases \(\phi\), and polarizations \(\mathbf{p(n)}\):

\begin{equation}
\label{one} 
\textbf B(\overrightarrow{r})=\sum_{i=0}^{N}A_{n}p_{n}\cos(\overrightarrow{\textbf r}\cdot\overrightarrow{\textbf k}_{n}+\phi_{n})
\end{equation}

The generation of \(A_{n}\), \(\phi\), and \(\mathbf{p(n)}\) was implemented using the Monte Carlo method [5]. The magnetic field obtained in this way is called "synthetic"; this approach for generating the magnetic field is a classical method in studies of cosmic ray diffusion in turbulent magnetic fields and is discussed in detail in reference [6].

The second part of the work involved calculating the spectra of synchrotron radiation from model particles in the synthetic magnetic fields obtained. As mentioned earlier, this study primarily focused on relativistic electrons. Observing spatial inhomogeneities in synchrotron radiation will allow for the search for inhomogeneous structures in the magnetic field of the galaxy and enable an analysis of the behavior of high-energy charged particles located in the detected inhomogeneous structures.

From Earth, synchrotron radiation can be observed in a limited frequency range from 10 MHz to 10 GHz. This range is due to absorption in regions of ionized interstellar hydrogen at the lower boundary and microwave background radiation at the upper boundary [2].
	
In this work, an analysis was conducted on the dependence of the shape of the total power spectrum \(P(\omega)\) on the variation of different values of the magnetic field strength \(\mathbf{B}\) and the pitch angle \(\alpha\) (the angle between the velocity vector of the charged particle and the direction of the magnetic field), with the aim of identifying the shift of the spectral peak of total power depending on the contributions from various spatial regions (Fig. 2).

\begin{figure*}
\centering{\includegraphics[scale=0.4]{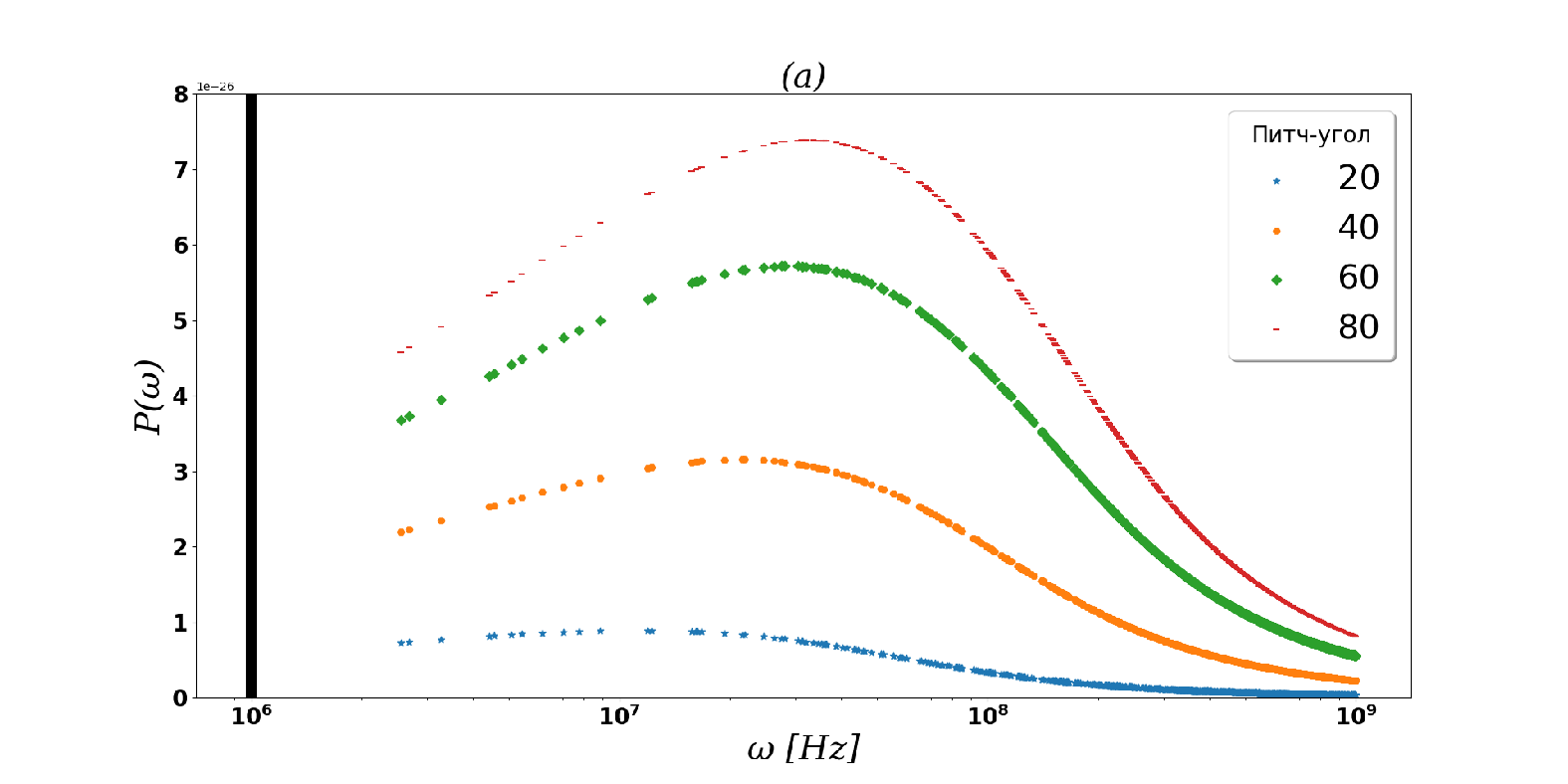}}
\centering{\includegraphics[scale=0.4]{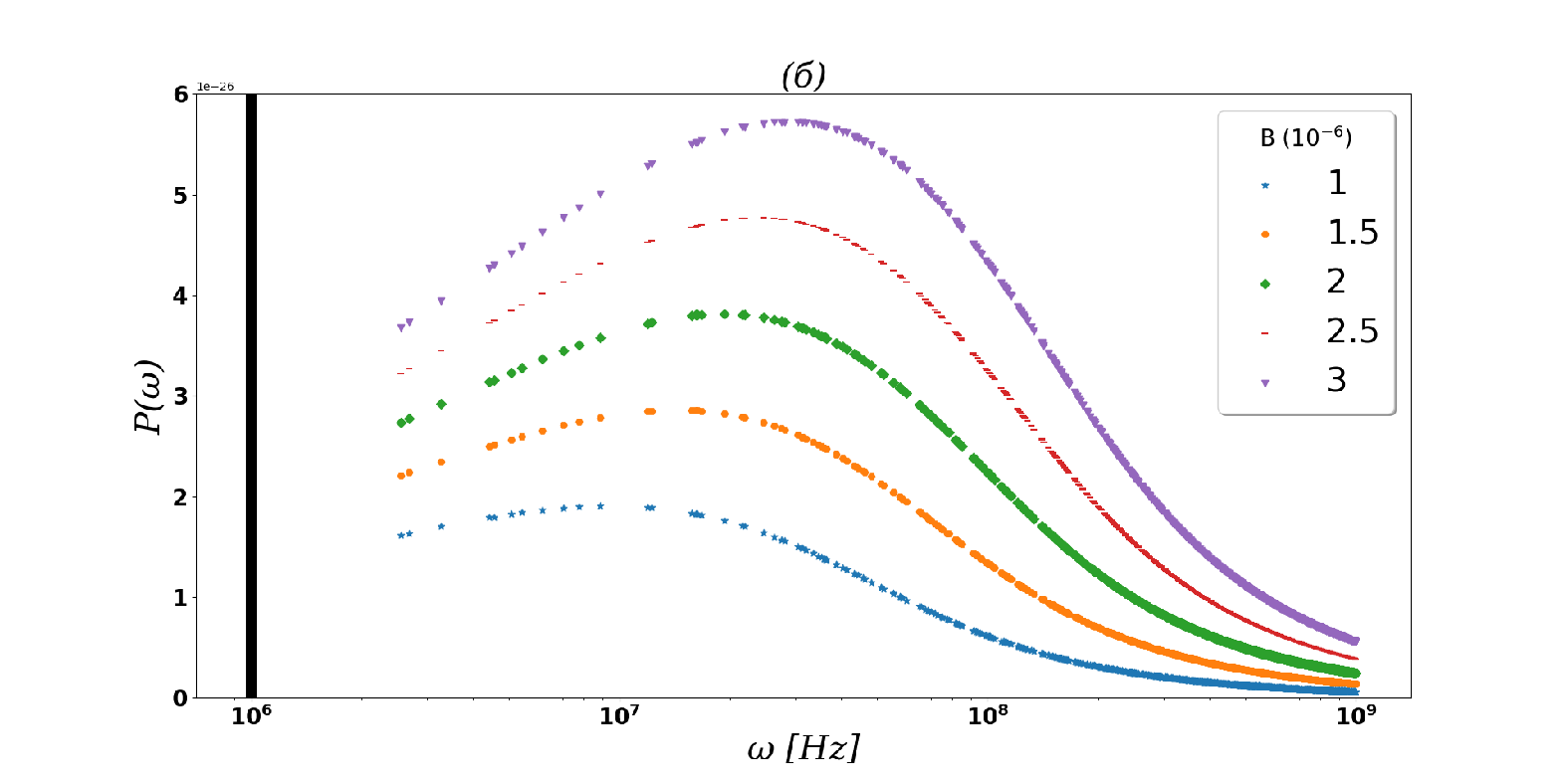}}
\caption{a) Spectra of total radiation power as a function of the pitch angle \(\alpha\); b) Spectra of total radiation power as a function of the magnetic field strength \(\mathbf{B}\); The black dashed line represents the boundary of the radio window. }
\end{figure*}

The constructed spectra demonstrate an increase in the total radiation power in the radio window with an increase in the pitch angle, as well as a noticeable shift with the growth of the local magnetic field strength at the emission point. This behavior of the spectrum provides insight into the contributions of local parameters of spatial inhomogeneity to the overall spectrum of total power.

Next, a model of the synchrotron background of the sky in the radio window was generated with an arbitrary reference point for angular coordinates, containing the total power spectra at arbitrary points in space within a synthetic magnetic field. The idea of this model is based on the isotropic generation of \(N\) point sources of synchrotron radiation. In each such source, an isotropic flux of relativistic electrons, normalized to unity, is postulated, taking into account the distribution of gamma factors following a power law. The calculation of synchrotron losses was performed considering the angle between the line of sight to the point and the direction of the magnetic field at that point in space, as well as the value of the magnetic field strength. The total power spectrum at each point was computed using (Equation 7), integrated over the gamma factors, accounting for the normalization energy coefficient \(\left(\frac{\gamma_{0}}{\gamma}\right)^{\delta}\). Here, \(\delta\) is the spectral index of electrons in the GC in the energy range \(\gamma_{\text{min}} = 1600\) and \(\gamma_{\text{max}} = 52000\), which corresponds to the energy range that makes a significant contribution to the radio window [2]. The spectral index was obtained by approximating experimental data from CRDB [7]:

\begin{equation}
\label{one} 
P(\omega)=f(B,\sin(\alpha))\int_{\gamma_{min}}^{\gamma_{max}}(\frac{\gamma_{0}}{\gamma})^{\delta}F(\frac{\omega}{\omega_{c}})d\gamma,
\end{equation}

where

\begin{equation}
\label{one} 
f(B,\sin(\alpha))=\frac{\sqrt{3}}{2\pi}\frac{e^{3}B\sin(\alpha)}{mc^{2}}
\end{equation}

Figure 3 shows an image of the projection of three-dimensional space onto a two-dimensional layout of the celestial sphere in the form of a histogram, where each bin contains the sum of the total powers of synchrotron radiation at arbitrary points in space.

\begin{figure*}
\centering{\includegraphics[scale=0.3]{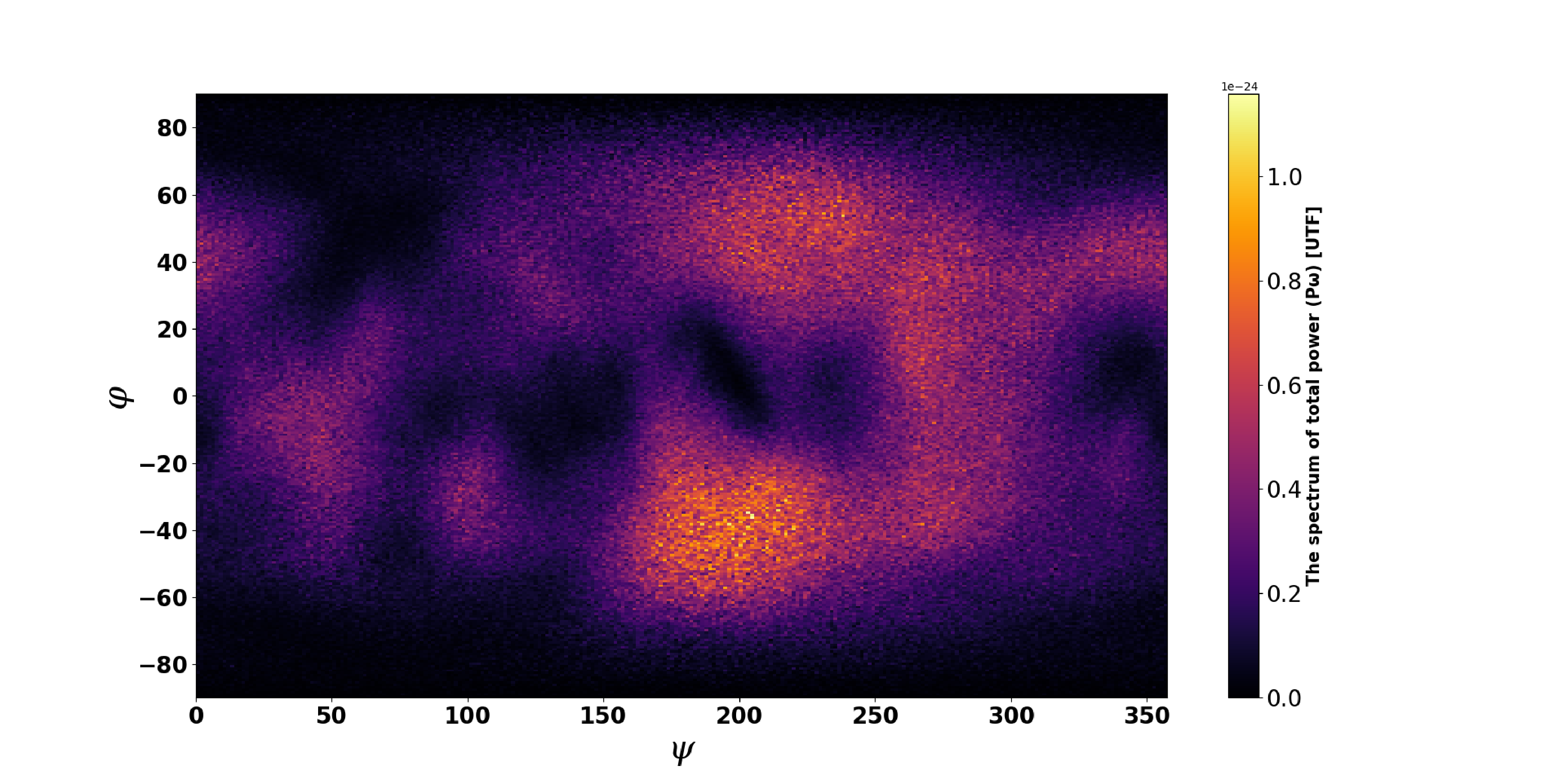}}
\caption{The calculated intensity of synchrotron losses of electrons in a synthetic magnetic field in polar coordinates, observed from the center of the generated volume.}
\label{G-CCD}
\end{figure*}

The image constructed within our model shows that there are noticeable structures with varying intensities of radiation in different regions of space. This is related to the characteristics of the magnetic field in these areas.

The final stage of the current work was the demonstration of the proposed method for calculating radiation and determining the characteristic sizes of the structure of spatial inhomogeneity. A model volume of space with a radius of 100 pc was generated, and the calculated radiation was projected onto a two-dimensional image of the sky, observed from the center of the sphere. The method for estimating the sizes of spatial inhomogeneity involved comparing local windows of different sizes on the obtained two-dimensional image. For an example image sized \(150 \times 150\) pixels, windows of sizes \(2 \times 2\), \(4 \times 4\), …, \(150 \times 150\) pixels were taken in all possible combinations across the constructed image. Then, the difference between the maximum and minimum power was calculated by summing all pixels in the window for each window variation. The result was a diagram that shows the pixel sizes of the inhomogeneous spatial structure (Fig. 4).

\section{Conclusion}

The authors have developed a methodology that allows for the calculation of the observed structure of spatial inhomogeneities in the synchrotron radiation of the GC in a turbulent magnetic field. In subsequent works, the method will be applied to construct a model of the sky of diffuse radio emission from the GC within the volume of the galaxy, taking into account the regular large-scale component of the magnetic field. The analysis of the final image will be conducted using both the method proposed in this work and the method of spherical harmonics (multipoles) and wavelet analysis. Such a model could serve as a useful tool for studying the distribution of magnetic fields and relativistic particles in space.

\begin{figure*}
\centering{\includegraphics[scale=0.4]{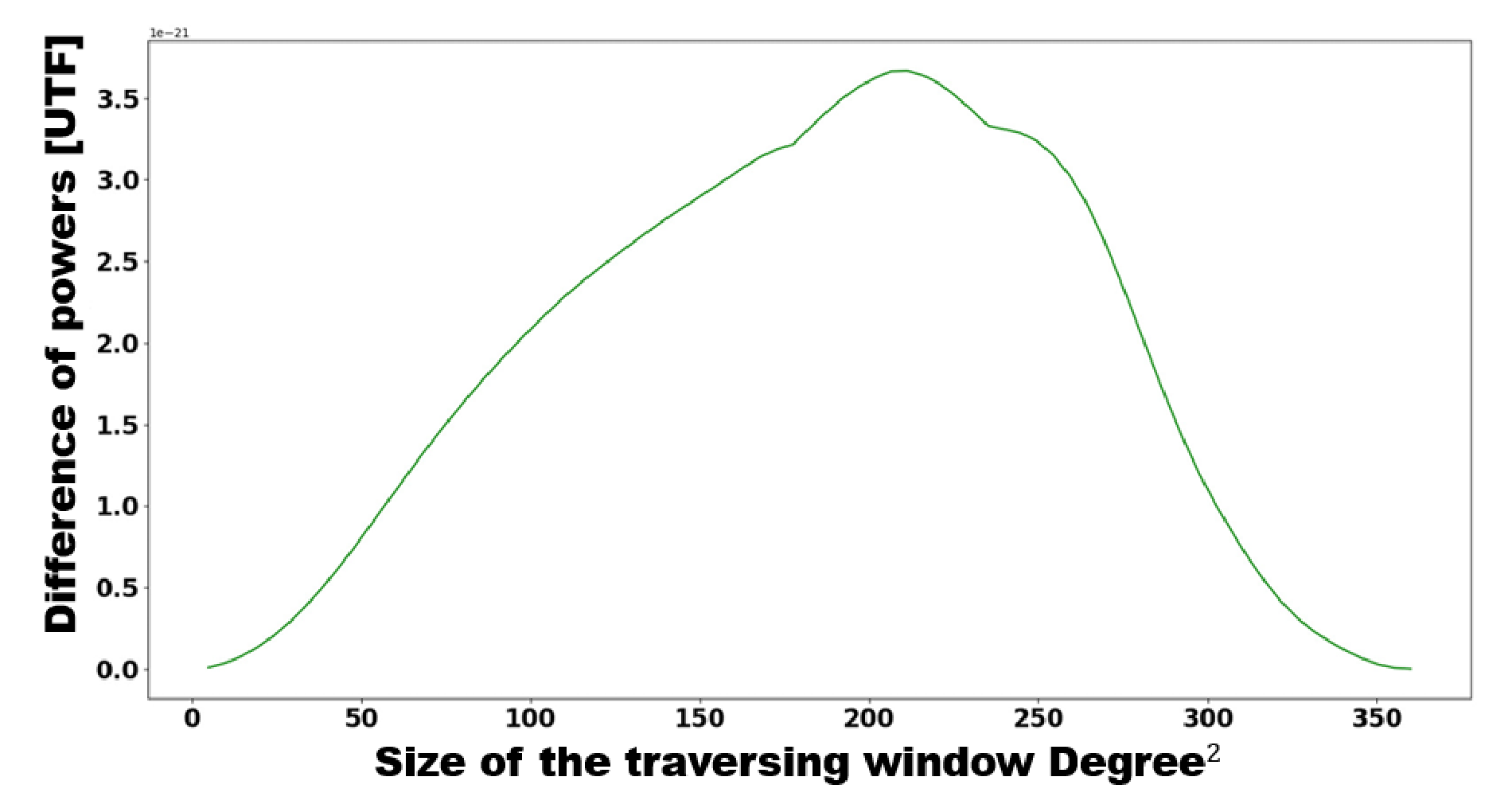}}
\caption{The dependence of the difference between the maximum and minimum total radiation power in the radio range, which is transparent for observation from Earth, shows that the behavior of the curve is independent of the pixel resolution of the window.}
\label{G-CCD}
\end{figure*}

\end{multicols}{}
\end{document}